\documentclass[conference]{IEEEtran}
\IEEEoverridecommandlockouts
\everymath{\displaystyle}
\usepackage{cite}
\usepackage{amsmath,amssymb,amsfonts}
\usepackage{algorithmic}
\usepackage{graphicx}
\usepackage{textcomp}
\usepackage[dvipsnames]{xcolor}
\usepackage{hyperref}
\hypersetup{hidelinks,colorlinks=false}
\def\BibTeX{{\rm B\kern-.05em{\sc i\kern-.025em b}\kern-.08em
    T\kern-.1667em\lower.7ex\hbox{E}\kern-.125emX}}
\begin{document}

\title{Intelligent Traffic Signal Control System by Using Image Information}

\author{\IEEEauthorblockN{1\textsuperscript{st} Zong-Ming Lin}
\IEEEauthorblockA{\textit{Department of Electrical Engineering} \\
\textit{National Taipei University}\\
SanXia, New Taipei City,\\Taiwan, R.O.C.}
\and
\IEEEauthorblockN{2\textsuperscript{nd} Cheng-Yang Chang}
\IEEEauthorblockA{\textit{Department of Electrical Engineering} \\
\textit{National Taipei University}\\
SanXia, New Taipei City,\\Taiwan, R.O.C.}
\and
\IEEEauthorblockN{3\textsuperscript{rd} Chin-Yu Hu}
\IEEEauthorblockA{\textit{Department of Electrical Engineering} \\
\textit{National Taipei University}\\
SanXia, New Taipei City,\\Taiwan, R.O.C.}
\and
\IEEEauthorblockN{4\textsuperscript{th} Yung-Yuan Chen}
\IEEEauthorblockA{\textit{Department of Electrical Engineering} \\
\textit{National Taipei University}\\
SanXia, New Taipei City,\\Taiwan, R.O.C.}
}

\maketitle

\begin{abstract}
This paper implements a traffic signal control system by using real-time traffic flow feedback. This system is designed to deal with two-lane intersections. We construct an experiment field similar to the roads and drivers in Taiwan using an autonomous simulation software called Virtual Test Drive (VTD) released by MSC Software \cite{b0}. We erect four cameras on the side of the roads to get the image of the intersection, then transfer the image information into traffic flow information. Analyze the traffic information in each lane by using Greenshields traffic flow model \cite{b1}. Control the traffic signals by using Webster's method \cite{b2} to increase the performance and soothe the traffic.
\end{abstract}


\section{Introduction}

The traditional traffic signal control system is based on the time and date, and it cannot respond to the road conditions in time. With the formation of the innovative city concept, the traditional traffic control system has been unable to cope with the ever-increasing traffic volume. According to the report in America, people wasted 20 percent of their travel time by waiting for the red lights \cite{b3}.

The road re-planning project is expensive, and it is pretty dangerous for the traffic police standing in the middle of the intersection to relieve the traffic. Intelligent traffic control systems become a decent way to soothe traffic. In this project, we use camera images as the system input to deal with the traffic problem to make the traffic signal control system more intelligent and sufficient, hoping Taiwan becoming a smart city in the future.

\section{Related Work}

Mukremin Ozkul published an article that proposed a safe traffic control model (STCM) based on anonymous vehicle messages to obtain driving information of vehicles approaching an intersection to ensure the authenticity of the information \cite{a1}. This article uses Simulation of Urban Mobility (SUMO), open-source software for traffic simulation, to simulate a single two-way four-lane intersection with a left turn signal. The experiment controls the number of vehicles passing through the simulation operation and compares it with the fixed-time signal control and Webster's signal control, proving that this method performs better with the traditional one.

Mengyu Guo used the Internet of Vehicles to obtain all vehicles' speed and position information and use Q-Learning in reinforcement learning to optimize traffic for controlling traffic lights at a single large intersection \cite{a2}. The paper's experiment used three other different algorithms to compare that the Q-Learning used in the paper can effectively alleviate the traffic flow.

At present, most intelligent signal systems are optimized for traffic at extensive and well-planned intersections in cities. Therefore, this project intends to use VTD software to improve the traffic of unplanned small and medium intersections.

\section{System Approach}
\subsection{Brief Introduction}

The intelligent traffic control system uses Webster's method as the controller to decide the phase time length of each road. The method needs the traffic flow ratio of each road, namely, the real-time traffic flow divided by the maximum traffic flow. Therefore, we only need to get two parameters from the intersection: ``maximum traffic flow'' and ``real-time traffic flow''. The maximum traffic flow is precalculated by Greensheids theory, whereas cameras and programs obtain the real-time traffic flow.

\subsection{Obtain the Maximum Traffic Flow}

The maximum traffic flow represents the maximum number of vehicles that can pass through in a time. Many factors can affect the maximum traffic flow, the lane width, the driver's driving style, the driver's visibility, and the type of the car. Therefore, we use the Greensheids traffic flow model to obtain the traffic model on our experiment track and get the maximum traffic flow parameter.

\begin{figure}[htbp]
    \centerline{\includegraphics[scale=0.4]{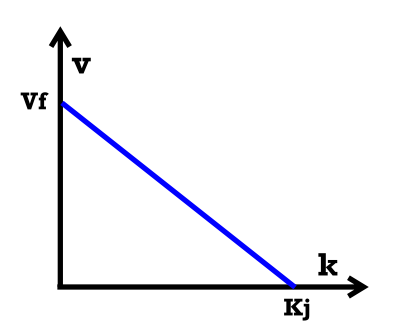}}
    \caption{Speed-Density Relation.}
    \label{v-k}
\end{figure}

\begin{figure}[htbp]
    \centerline{\includegraphics[scale=0.4]{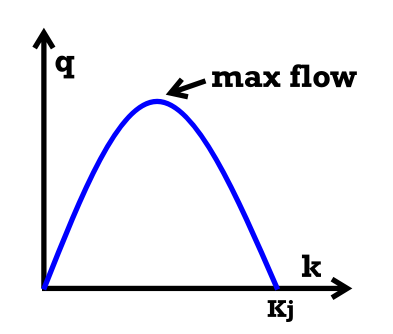}}
    \caption{Flow-Density Relation.}
    \label{q-k}
\end{figure}

Greenshields flow model describes the behavior of the vehicles on the roads, the relation between the vehicle speed, vehicle density, and vehicle flow. The model has three functions: ``speed-density relation'' in Fig.~\ref{v-k}, ``definition of vehicle flow'' in \eqref{flow-eq}, and the ``flow-density relation'' in Fig.~\ref{q-k}.

\begin{equation}
    Flow = Speed \times Density \label{flow-eq}
\end{equation}

The function in Fig.~\ref{v-k} of the Greenshields flow model can also be read as the relation between the vehicle speed and the distance between two consecutive vehicles. The faster the driver drives, the safety distance with the car in front increases. Therefore, when the road with the same distance has a smaller capacity for the vehicles, it also represents the density of the road gets lower than usual.

\begin{figure}[htbp]
    \centerline{\includegraphics[scale=0.4]{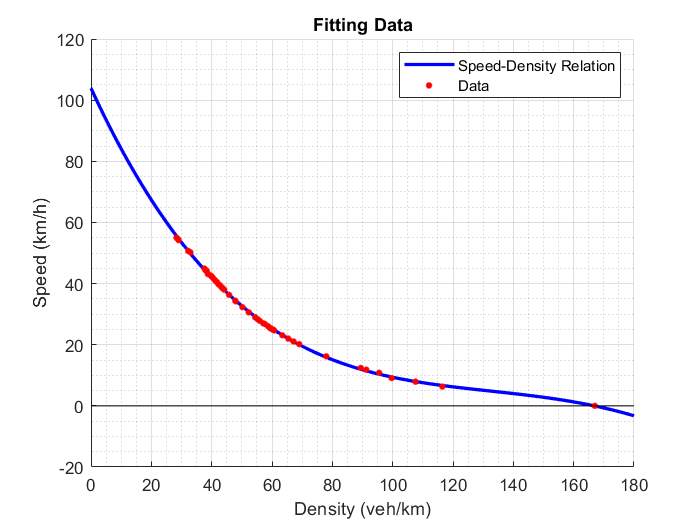}}
    \caption{Custom Speed-Density Relation.}
    \label{custom v-k}
\end{figure}

\begin{figure}[htbp]
    \centerline{\includegraphics[scale=0.4]{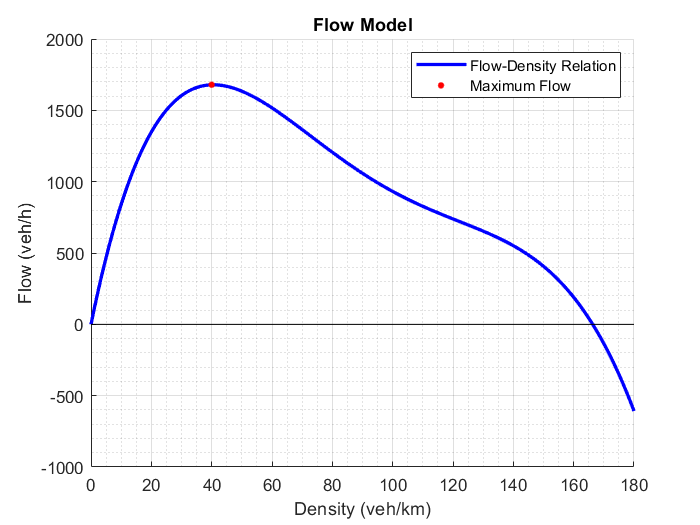}}
    \caption{Custom Flow-Density Relation.}
    \label{custom q-k}
\end{figure}

We collected the vehicles' speed and distance data in our simulation map. Next, we get our speed-density relation in Fig.~\ref{custom v-k} by using the curve fitting method with the MATLAB toolbox. Then, substitute the definition of vehicle flow \eqref{flow-eq} to obtain the flow-density relation in Fig.~\ref{custom q-k}. Finally, we can get the maximum flow by finding the maximum point from the flow-density relation we have made, which is 1679 vehicle passes per hour.

\subsection{Obtain the Real-time Traffic Flow}

The real-time traffic flow means how many vehicles pass through in a period. There are two parts to this section. The first part is the vehicle counter, and the second part is to decide the time length of calculating a result.

\begin{figure}[htbp]
    \centerline{\includegraphics[scale=0.15]{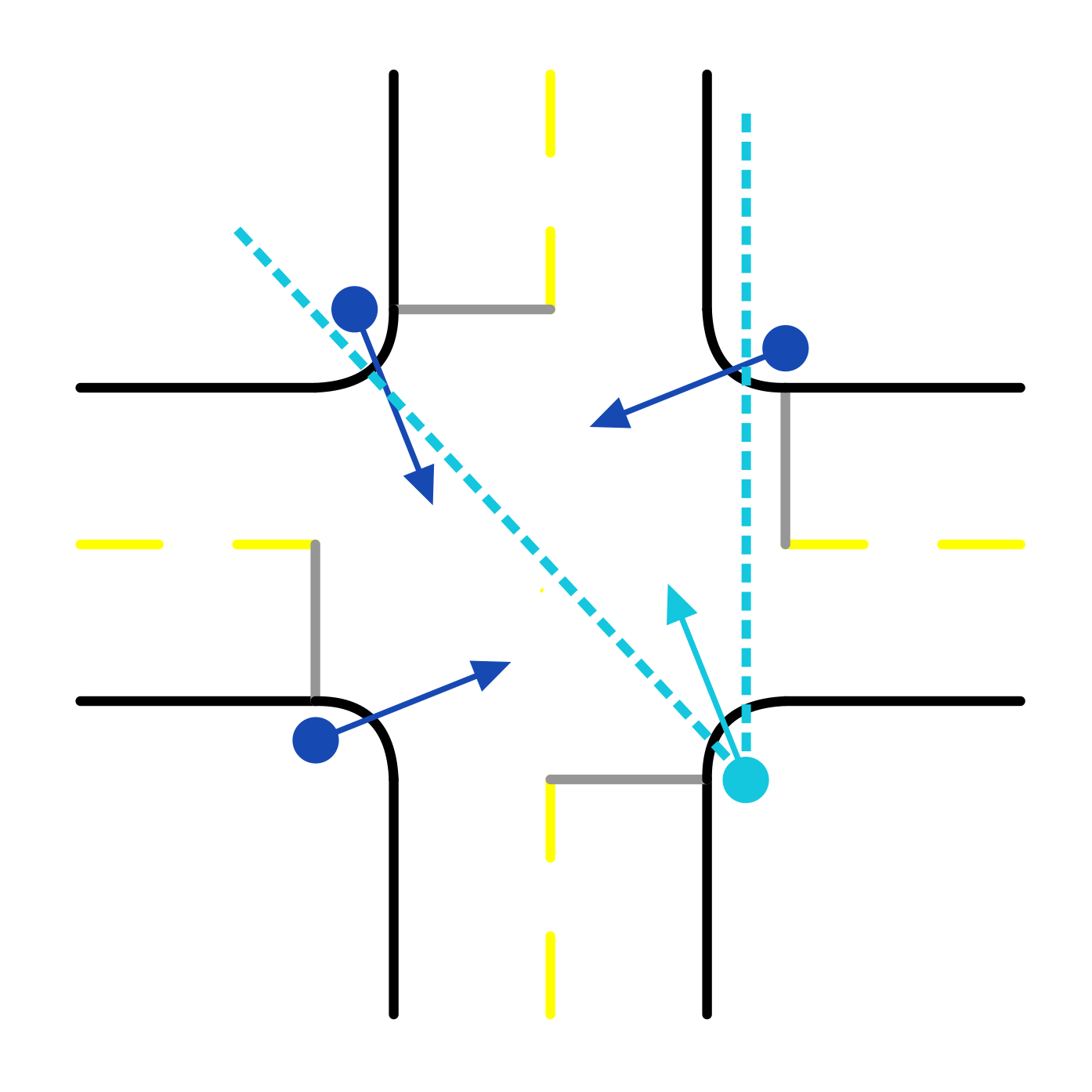}}
    \caption{Positions of the Four Cameras.}
    \label{cam pos}
\end{figure}

We use four cameras erect on the lamppost, which is near the intersection, to watch out for the opposite side of the traffic conditions shown in Fig.~\ref{cam pos}.

\begin{figure}[htbp]
    \centerline{\includegraphics[scale=0.35]{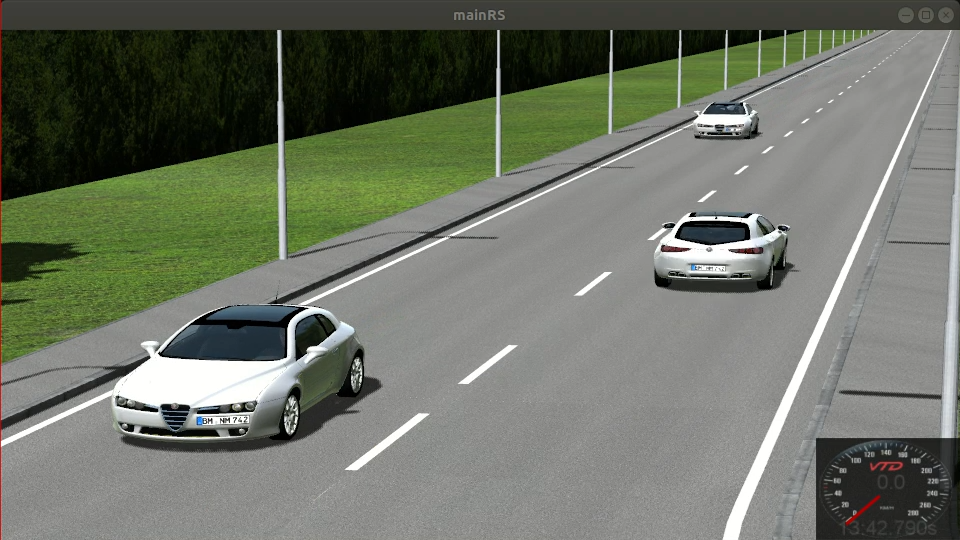}}
    \caption{The Veiw of the Camera.}
    \label{cam view}
\end{figure}

\begin{figure}[htbp]
    \centerline{\includegraphics[scale=0.35]{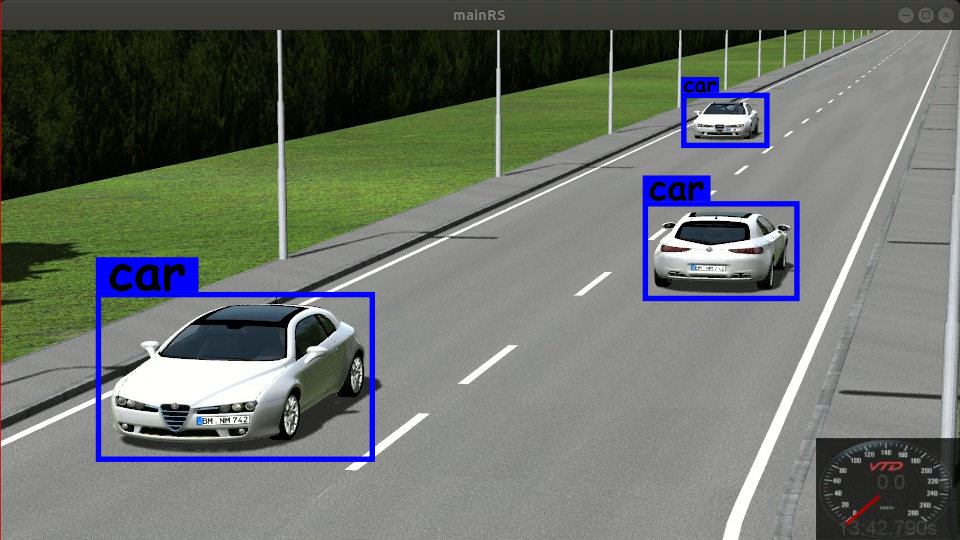}}
    \caption{The Result of Object Detection.}
    \label{object detection}
\end{figure}

\begin{figure}[htbp]
    \centerline{\includegraphics[scale=0.35]{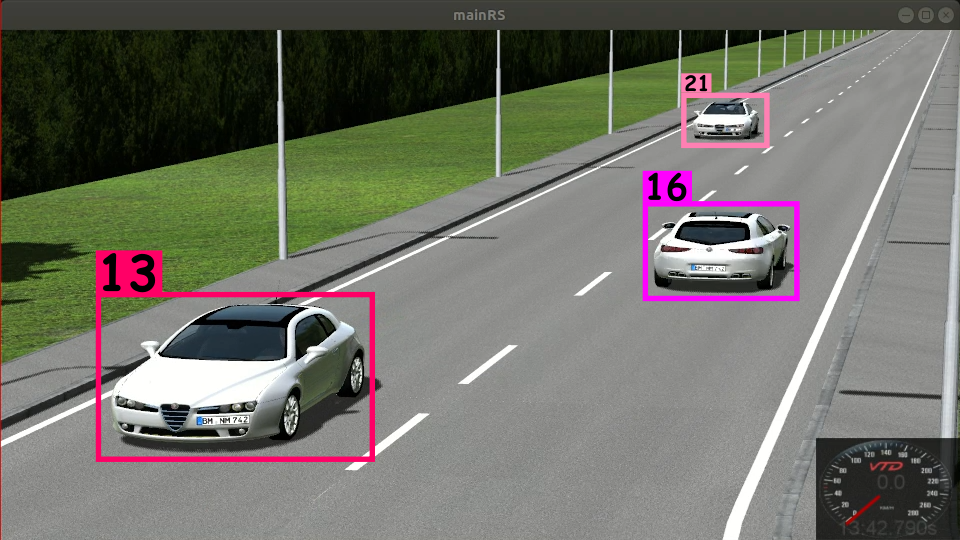}}
    \caption{The Result of Multiple Object Tracking.}
    \label{MOT}
\end{figure}

The image obtained from the cameras is in Fig.~\ref{cam view}. First, we use YOLOv4-Tiny \cite{b4}, a quick machine learning method that suits embedded systems, to detect the cars in the images where an example is shown in Fig.~\ref{object detection}. Next, we use DeepSORT \cite{b5}, a multiple object tracking method using Kalman filter and Hungarian algorithm, which gives every single vehicle a specific identification number. The result is shown in Fig.~\ref{MOT}.

By using the object tracking information. We use an easy algorithm to count the number of vehicles that enter the intersection, namely, the left-hand lane in the image.

\begin{figure}[htbp]
    \centerline{\includegraphics[scale=0.35]{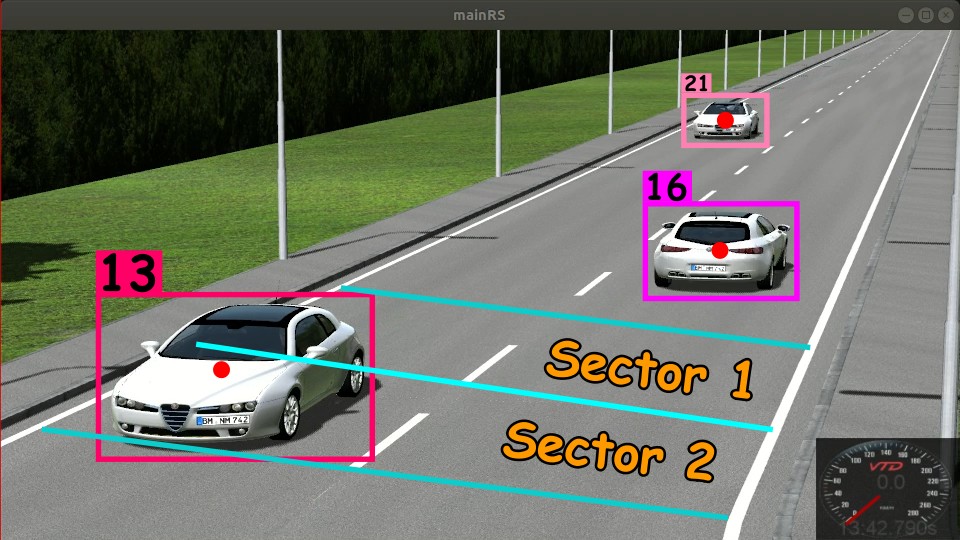}}
    \caption{Vehicle Counting Method.}
    \label{counting}
\end{figure}

Divide the road into two sections shown in Fig.~\ref{counting}. Use the center of the bounding box as the representative of the vehicle. The main idea of the algorithm is that if the center moves from sector 1 to sector 2, then the counter of the lane increases by one. There are two minor changes to enhance the performance of the counter. We record the car's ID that successfully increased the counter to prevent recalculating the same car again. Also, we keep the ID that has been in sector 1 for one second in case the bounding box somehow disappears for a short time and reappear in sector 2. With the camera frame rate at 5 frames per second, the accuracy of the vehicle counter is between 96.8$\sim$97.4\%, which is reliable enough for the intelligent traffic signal control system.

\begin{figure}[htbp]
    \centerline{\includegraphics[scale=0.18]{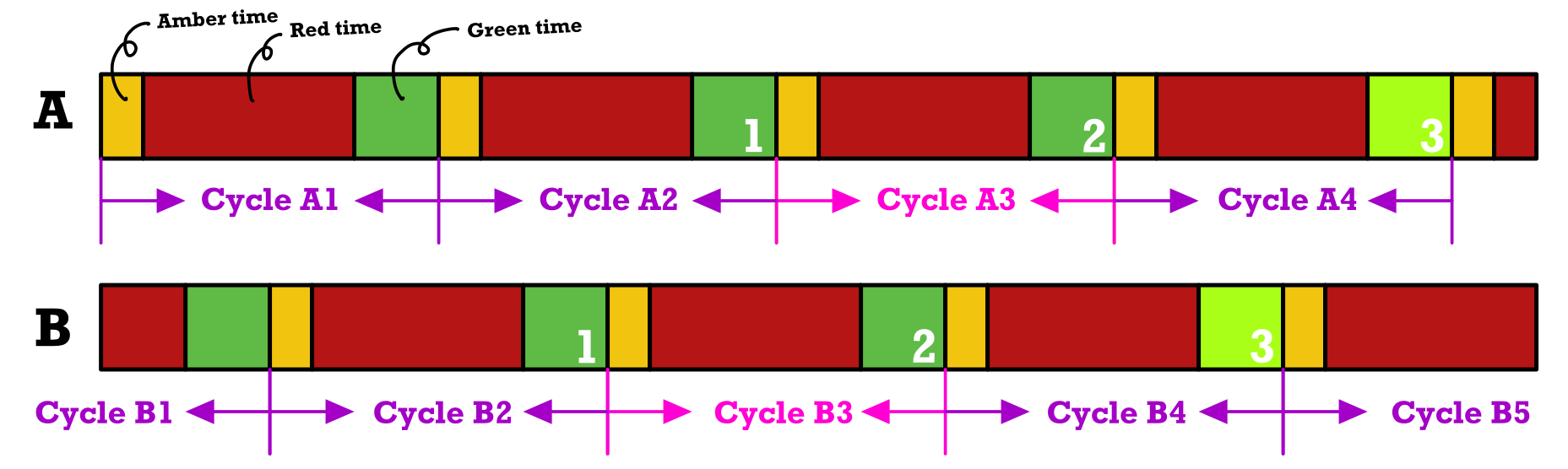}}
    \caption{The Definition of a Time Length.}
    \label{time length}
\end{figure}

As for deciding the time length of each calculation, we decided that every cycle time length is started from the amber time begins and stops when the green time ends, as shown in Fig.~\ref{time length}. The reason for the decision is that if the time length is too short, the vehicle flow could be affected by the signal lights; if the time length is too long, it loses the meaning of real-time control.

Assume the vehicle flow is like water flowing through a river, and the traffic lights are like a dam across the river. When the dam is opened, the water constantly flows through the river. When the dam is closed, the water is hoarded by the dam. When the dam reopened again, the water drained at a faster rate. If the river's flow rate recurs the original constant flow, then it means that the flow rate can be calculated as the water volume that flowed through the dam from the time the dam closed to the dam reclosed again. Therefore, transferring the idea to traffic flows, the calculating time length is from the green phase turns to amber the next time the green phase ends.

We can be able to count the vehicles, and we also decided the calculating time length. Thus, the real-time vehicle flow can be obtained as \eqref{real-time flow}.

\begin{equation}
    \textit{Real-time vehicle flow} = \frac{\textit{The number vehicles passes by}}{\textit{Defined time length}} \label{real-time flow}
\end{equation}

\subsection{Phase Time Computation}

We use Webster's method as the phase time distribution. The inputs are the vehicle flow ratio of each road, and the outputs are the time of green phase for each road.

There are two directions lanes on each road. Every lane has its camera and produces a real-time vehicle flow. Instead of summing the two flow data up, we chose to pick the maximum value among the two data as the function input. We think that doing it in this way can reflect the need of the road.

\begin{figure}[htbp]
    \centerline{\includegraphics[scale=0.35]{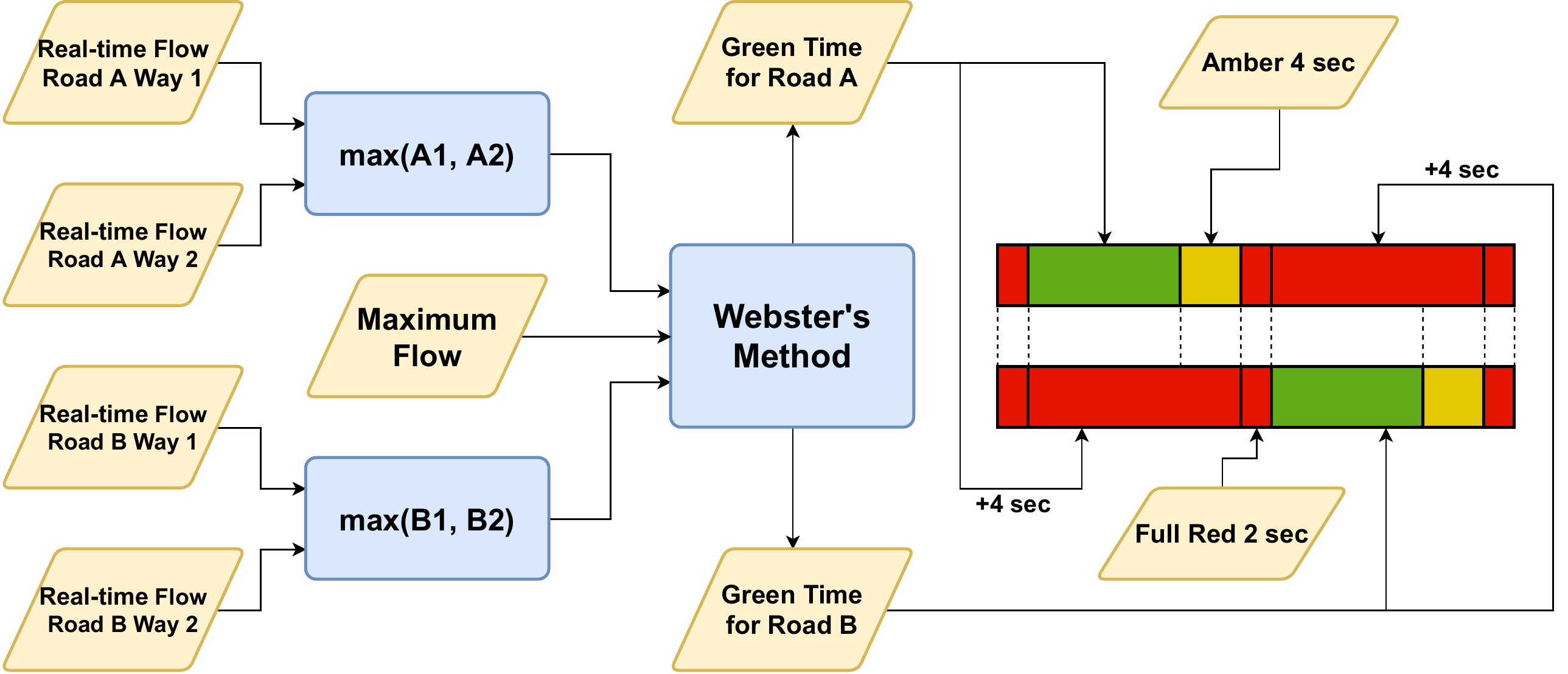}}
    \caption{Phase Time Computation.}
    \label{phase time}
\end{figure}

Because the amber time is 4 seconds and the all-red time is 2 seconds which were all defined by the traffic rules \cite{b6}. Moreover, if one of the roads is at the green phase, the diagonal road must be at the red phase. Therefore, we only need to obtain the time of the green phase for each road; then, we can expand it into the complete phase schedule shown in Fig.~\ref{phase time}.

\begin{figure}[htbp]
    \centerline{\includegraphics[scale=0.1]{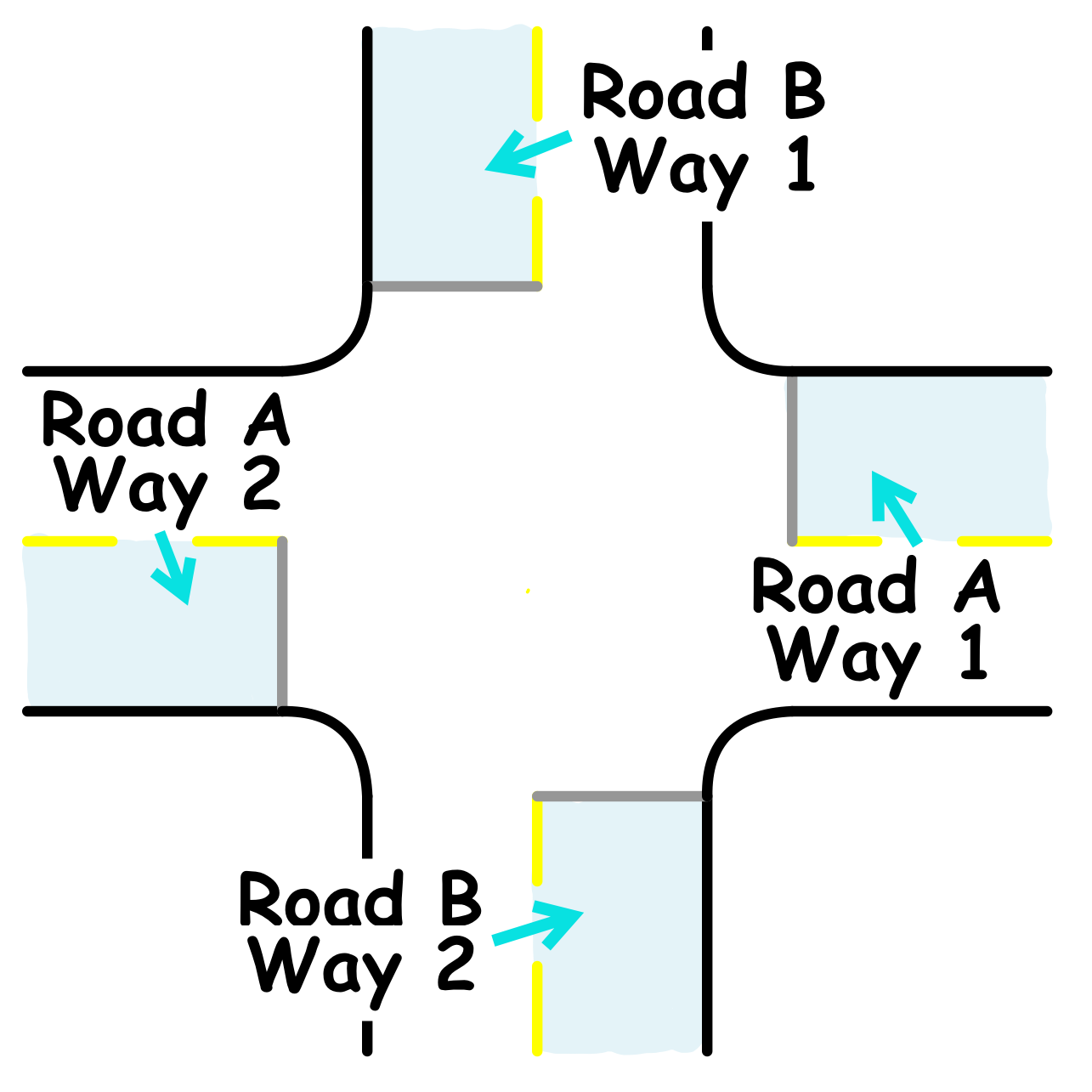}}
    \caption{The Codename of Each Lane.}
    \label{road number}
\end{figure}

Green phase time is calculated by Webster's method as below:

\begin{align}
    \textit{Flow ratio} \Rightarrow \begin{cases}
        \textit{Road A flow ratio}\:y_a &= \frac{\textit{Real-time flow}_A}{\textit{Maximum flow}} \\
        \textit{Road B flow ratio}\:y_b &= \frac{\textit{Real-time flow}_B}{\textit{Maximum flow}} \\
    \end{cases} \label{flow ratio}
\end{align}
\begin{align}
    \textit{Sum of flow ratio}\:Y &= y_a + y_b \\  \label{Y}
    \textit{Lost time}\:L &= \textit{Phase number} \nonumber\\ 
    &\times (\textit{Amber time} + \textit{All red time}) \\
    \textit{Ideal cycle time}\:C_0 &= \frac{1.5 \cdot L}{|1-Y|} \label{ideal}
\end{align}
\begin{align}
    &\textit{Green phase time} \Rightarrow \nonumber\\
    &\begin{cases}
        \textit{Green time for road A}\:G_a &= \frac{y_a}{Y} \cdot |C_0-L| \\
        \textit{Green time for road B}\:G_b &= \frac{y_b}{Y} \cdot |C_0-L| \\
    \end{cases}
\end{align}
\begin{align}
    \textit{Single cycle for the signal control} &= \nonumber\\
    G_a + G_b + 2 \times ( \textit{Amber}\:4\:sec &+ \textit{All red}\:2\:sec )
\end{align}

Noted that the "phase time" parameter is 2 because there are two green lights ($G_a,\:G_b$) in total. Moreover, the absolute value in those formulas is added by us additionally since sometimes the real-time vehicle flow \eqref{flow ratio} change may let the sum of the flow ratio value \eqref{Y} be greater than 1, causing the Ideal cycle time \eqref{ideal} less than 1, and letting the green phase time become a minus number. The sum of flow ratio could not be greater than 1. If so, then it means the road is overloaded and cannot relieve all the vehicles on the road even though the traffic signal is optimized in the most significant way.

\section{Simulation Experiment}
\subsection{Map Settings}

We use an autonomous vehicle simulation software called Virtual Test Drive (VTD) to test whether the intelligent traffic signal control system that we have designed can improve the traffic or not. We create our map to experiment by using Road Designer (ROD). Setup the traffic lights and driver parameters by using Scenario Editor.

\begin{figure}[htbp]
    \centerline{\includegraphics[scale=0.15]{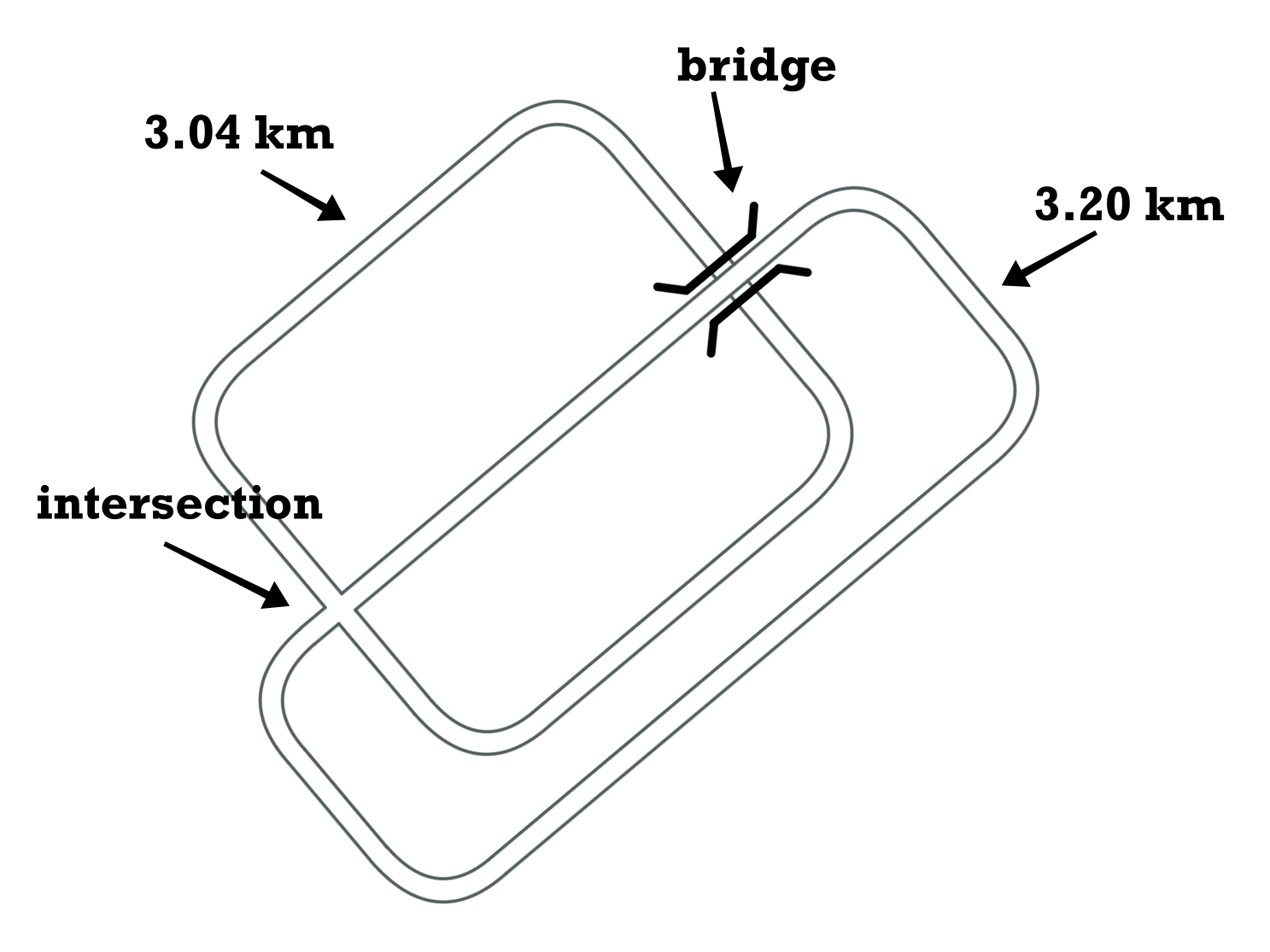}}
    \caption{The Map was Used for the Experiment.}
    \label{map}
\end{figure}

\begin{figure}[htbp]
    \centerline{\includegraphics[scale=0.18]{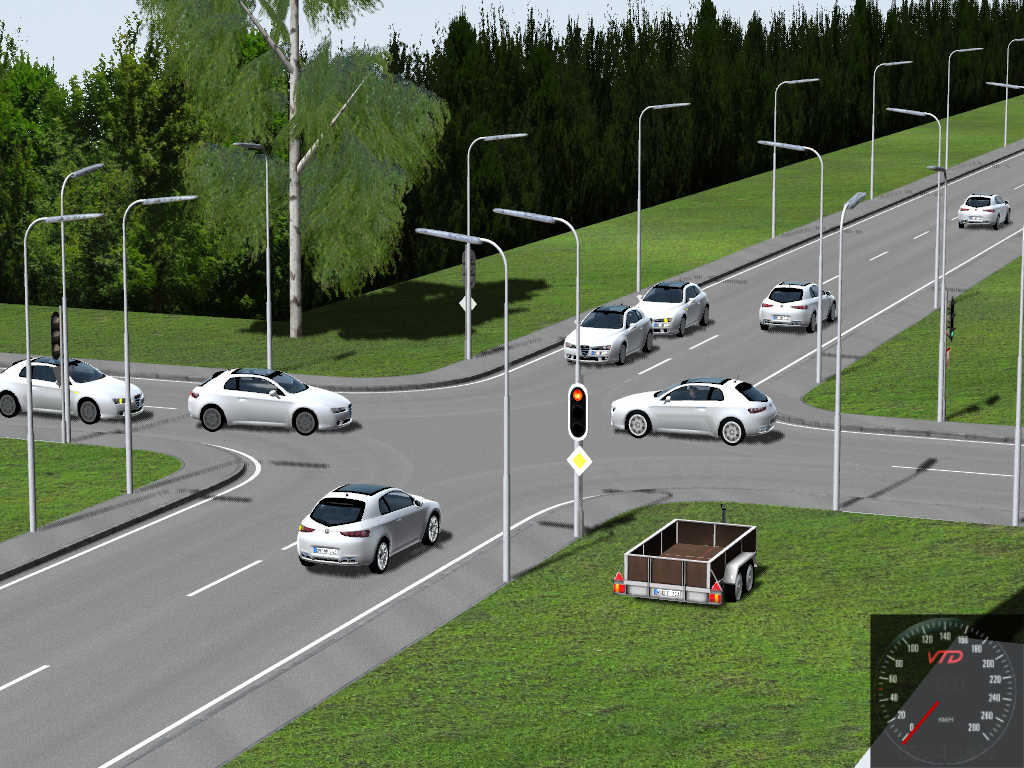}}
    \caption{A view of the experiment space.}
    \label{secne}
\end{figure}

The map has two rectangular rings representing the two roads of an intersection perpendicular to each other. The left ring has a length of 3.04 kilometers, whereas the right ring has 3.20 kilometers. To measure the performance of the traffic control system, we only want one intersection rather than two. Therefore, we built a bridge to staggered one of the intersections, as shown in Fig.~\ref{map}. Moreover, we decorated the experiment space with Street lamps, grass, and trees to make it more exquisite, Fig.~\ref{secne}.

The only vehicle type we used for the experiment is the Alfa Romeo Giulia which vehicle length is 4.643 meters. We chose the insecure driver in the simulation software option as the driving style of the vehicle drivers. This option is the most similar to the drivers in Taiwan. Although it is called an insecure driver, it drives completely legal and also has some characteristics while driving:

\begin{itemize}
    \item All of the drivers have the same behavior.
    \item Obey traffic rules.
    \item Do not exceed speed limit of 60 km/h.
    \item Only slows down in the intersection.
    \item Do not slow down in turns.
    \item Do not exceed track limits.
    \item Do not overtake others.
\end{itemize}

The above characteristics make the vehicles only act differently due to the traffic light phase and having the same actions in the road sections.

We did two experiments to test its performance to measure the intelligent traffic signal control in different traffic situations. Experiment one tests the signal control in different traffic flows, whereas experiment two simulates the major/minor road traffic base on group three in the first experiment. According to the relation between traffic flow and vehicle density which we have made in Fig.~\ref{custom q-k}, put the appropriate number of vehicles in the experiment map to produce the corresponding traffic flows. The data of the groups are shown in TABLE~\ref{sym exp group} and TABLE~\ref{asym exp group}.

\begin{table}[htbp]
    \caption{Vehicle Amount in the Experiment Space in Experiment One.}
    \begin{center}
        \begin{tabular}{|c|c|c|c|}
            \hline
             & \textbf{Flow (veh/s)} & \textbf{Vehicle Amount} & \textbf{Left Amt. / Right Amt.}\\\hline
            \textbf{1} & 0.100 & 46 & 22 / 24 \\\hline
            \textbf{2} & 0.125 & 60 & 30 / 30 \\\hline
            \textbf{3} & 0.150 & 72 & 35 / 37 \\\hline
            \textbf{4} & 0.175 & 87 & 42 / 45 \\\hline
        \end{tabular}
        \label{sym exp group}
    \end{center}
\end{table}

\begin{table}[htbp]
    \caption{Vehicle Amount in the Experiment Space in Experiment Two.}
    \begin{center}
        \begin{tabular}{|c|c|c|c|}
            \hline
             & \textbf{Flow Ratio} & \textbf{Vehicle Amount} & \textbf{Left Amt. / Right Amt.}\\\hline
            \textbf{1} & 50 / 50 & 72 & 35 / 37 \\\hline
            \textbf{2} & 60 / 40 & 74 & 44 / 30 \\\hline
            \textbf{3} & 70 / 30 & 75 & 53 / 22 \\\hline
            \textbf{4} & 80 / 20 & 77 & 63 / 14 \\\hline
        \end{tabular}
        \label{asym exp group}
    \end{center}
\end{table}

The traffic light setup conforms to the traffic regulations in Taiwan. The amber phase lasts 4 seconds after the green phase ends. Also, before the next green phase begins, there is a 2-second red phase for all the traffic lights at the intersection.

\subsection{Benchmark}

To measure the performance of the intelligent traffic control system, we use the traditional fixed-time traffic control as the control group. The phase time of the fixed-phase traffic control is fixed and does not change during the experiment. Also, Webster's method calculated the time length of each phase according to the initial vehicle amount on each road. The calculation result is shown in TABLE~\ref{sym ini time} and TABLE~\ref{asym ini time}.

\begin{table}[htbp]
    \caption{Green Time for Experiment One of the Fixed-time Signal Control.}
    \begin{center}
        \begin{tabular}{|c|c|c|c|}
            \hline
             \textbf{Flow (veh/s)} & \textbf{Left Ring} & \textbf{Right Ring} & \textbf{Single Cycle}\\\hline
            \textbf{0.100} & 13.7 sec & 14.1 sec & 39.8 sec \\\hline
            \textbf{0.125} & 19.3 sec & 18.4 sec & 49.7 sec \\\hline
            \textbf{0.150} & 25.3 sec & 25.4 sec & 62.7 sec \\\hline
            \textbf{0.175} & 38.8 sec & 39.4 sec & 90.2 sec \\\hline
        \end{tabular}
        \label{sym ini time}
    \end{center}
\end{table}

\begin{table}[htbp]
    \caption{Green Time for Experiment Two of the Fixed-time Signal Control.}
    \begin{center}
        \begin{tabular}{|c|c|c|c|}
            \hline
             \textbf{Flow Ratio} & \textbf{Left Ring} & \textbf{Right Ring} & \textbf{Single Cycle}\\\hline
            \textbf{50 / 50} & 25.3 sec & 25.4 sec & 62.7 sec \\\hline
            \textbf{60 / 40} & 31.6 sec & 21.6 sec & 65.2 sec \\\hline
            \textbf{70 / 30} & 36.9 sec & 16.3 sec & 65.2 sec \\\hline
            \textbf{80 / 20} & 42.1 sec & 10.6 sec & 64.7 sec \\\hline
        \end{tabular}
        \label{asym ini time}
    \end{center}
\end{table}

The test group is the intelligent traffic signal control system, or ``smart signal control'' for short. Its phase time length is according to the real-time traffic flow but also uses the fixed-time signal control phase time for the initial cycle as shown in TABLE~\ref{sym ini time} and TABLE~\ref{asym ini time}.

\subsection{Result Computation}

We measured the system's performance by comparing the average speed of all the vehicles during the simulation. The speed data is obtained from the RDBViewer, a subprogram in VTD that can record the discrete speed data from all vehicles. \eqref{avgspd} is the formula that calculates the average speed of every group, and the calculation result is shown in TABLE~\ref{sym avg speed} and TABLE~\ref{asym avg speed}.

\begin{align}
    \textit{Average speed} = \sum_{i=1}^{\textit{vehicles}} \sum_{j=1}^{\textit{data}} \frac{\textit{speed data}_{ij}}{\textit{vehicles}\times \textit{data}} \label{avgspd}
\end{align}

\begin{table}[htbp]
    \caption{The Average Speed in Experiment One.}
    \begin{center}
        \begin{tabular}{|c|c|c|c|c|}
            \hline
              & \multicolumn{2}{c}{\textbf{Fixed-time Signal Control}} & \multicolumn{2}{|c|}{\textbf{Intelligent Signal Control}} \\\hline
            \textbf{Flow}  & \textbf{Average} & \textbf{Simulation} & \textbf{Average} & \textbf{Simulation} \\
            \textbf{(veh/s)}  & \textbf{Speed} & \textbf{Time} & \textbf{Speed} & \textbf{Time} \\\hline
            \textbf{0.100} & 56.322 kph & 15.2 min & 56.046 kph & 11.7 min \\\hline
            \textbf{0.125} & 54.226 kph & 15.2 min & 54.898 kph & 10.8 min \\\hline
            \textbf{0.150} & 53.704 kph & 15.2 min & 54.551 kph & 10.7 min \\\hline
            \textbf{0.175} & 48.716 kph & 15.2 min & 51.303 kph & 10.8 min \\\hline
        \end{tabular}
        \label{sym avg speed}
    \end{center}
\end{table}

\begin{table}[htbp]
    \caption{The Average Speed in Experiment Two.}
    \begin{center}
        \begin{tabular}{|c|c|c|c|c|}
            \hline
              & \multicolumn{2}{c}{\textbf{Fixed-time Signal Control}} & \multicolumn{2}{|c|}{\textbf{Intelligent Signal Control}} \\\hline
            \textbf{Flow}  & \textbf{Average} & \textbf{Simulation} & \textbf{Average} & \textbf{Simulation} \\
            \textbf{Ratio}  & \textbf{Speed} & \textbf{Time} & \textbf{Speed} & \textbf{Time} \\\hline
            \textbf{50 / 50} & 53.704 kph & 15.0 min & 54.551 kph & 10.7 min \\\hline
            \textbf{60 / 40} & 51.178 kph & 15.0 min & 51.267 kph & 11.4 min \\\hline
            \textbf{70 / 30} & 46.198 kph & 15.0 min & 50.518 kph & 11.1 min \\\hline
            \textbf{80 / 20} & 40.953 kph & 15.0 min & 48.007 kph & 11.0 min \\\hline
        \end{tabular}
        \label{asym avg speed}
    \end{center}
\end{table}

However, the result cannot express the performance very well. Although vehicles only slow down by the intersection, because of the environment, vehicles do not disappear but reenter the intersection again, so we don't know how many times the car slows down. Therefore, we convert the average speed results into the ``average time lost'' as a single-vehicle passes the intersection.

The first step is to calculate the ``total time lost'' shown as \eqref{ttl}. As we know, the car only slows down by the intersection. Therefore, we can calculate how much time is wasted by the traffic lights during the simulation. Next, we calculate the average number of times the vehicles pass the intersection \eqref{avg inter pass}. We define the cycle time as 3 minutes, which is the time cost when a vehicle leaves the intersection and reenters it again. Finally, we can obtain the average time lost with \eqref{avg time lost} by using the total time lost and the average intersection pass. The result after the following calculation is shown in TABLE \ref{sym final result} and TABLE \ref{asym final result}.

\begin{align}
    \textit{Total time lost} &= \frac{\textit{speed limit} - \textit{avg. speed}}{\textit{speed limit}} \times \textit{sim. time} \label{ttl} \\
    \textit{Average pass} &= \frac{\textit{sim. time} - \textit{total time lost}}{\textit{cycle time}} \label{avg inter pass}\\
    \textit{Average time lost} &= \frac{\textit{total time lost}}{\textit{avg. intersection pass}} \label{avg time lost}
\end{align}

\begin{table}[htbp]
    \caption{The Average Lost Time in Experiment One.}
    \begin{center}
        \begin{tabular}{|c|c|c|}
            \hline
            \textbf{Flow} & \textbf{Fixed-time Signal Control} & \textbf{Intelligent Signal Control} \\
            \textbf{(veh/s)} & \textbf{(sec)} & \textbf{(sec)} \\\hline
            \textbf{0.100} & 11.797 & 12.701 \\\hline
            \textbf{0.125} & 19.211 & 16.732 \\\hline
            \textbf{0.150} & 21.147 & 17.980 \\\hline
            \textbf{0.175} & 41.742 & 30.517 \\\hline
        \end{tabular}
        \label{sym final result}
    \end{center}
\end{table}

\begin{table}[htbp]
    \caption{The Average Lost Time in Experiment Two.}
    \begin{center}
        \begin{tabular}{|c|c|c|}
            \hline
            \textbf{Flow} & \textbf{Fixed-time Signal Control} & \textbf{Intelligent Signal Control} \\
            \textbf{Ratio} & \textbf{(sec)} & \textbf{(sec)} \\\hline
            \textbf{50/50} & 21.147 & 17.980 \\\hline
            \textbf{60/40} & 31.027 & 30.663 \\\hline
            \textbf{70/30} & 53.774 & 33.784 \\\hline
            \textbf{80/20} & 83.716 & 44.982 \\\hline
        \end{tabular}
        \label{asym final result}
    \end{center}
\end{table}

The average lost time is not the time that the car is waiting for the green lights. It is the time that starts to count from the vehicle decelerates and ends when the car accelerates back to the speed limit. The time lost is the additional time needed comparing to the situation in which the intersection does not exist.

\begin{figure}[htbp]
    \centerline{\includegraphics[scale=0.4]{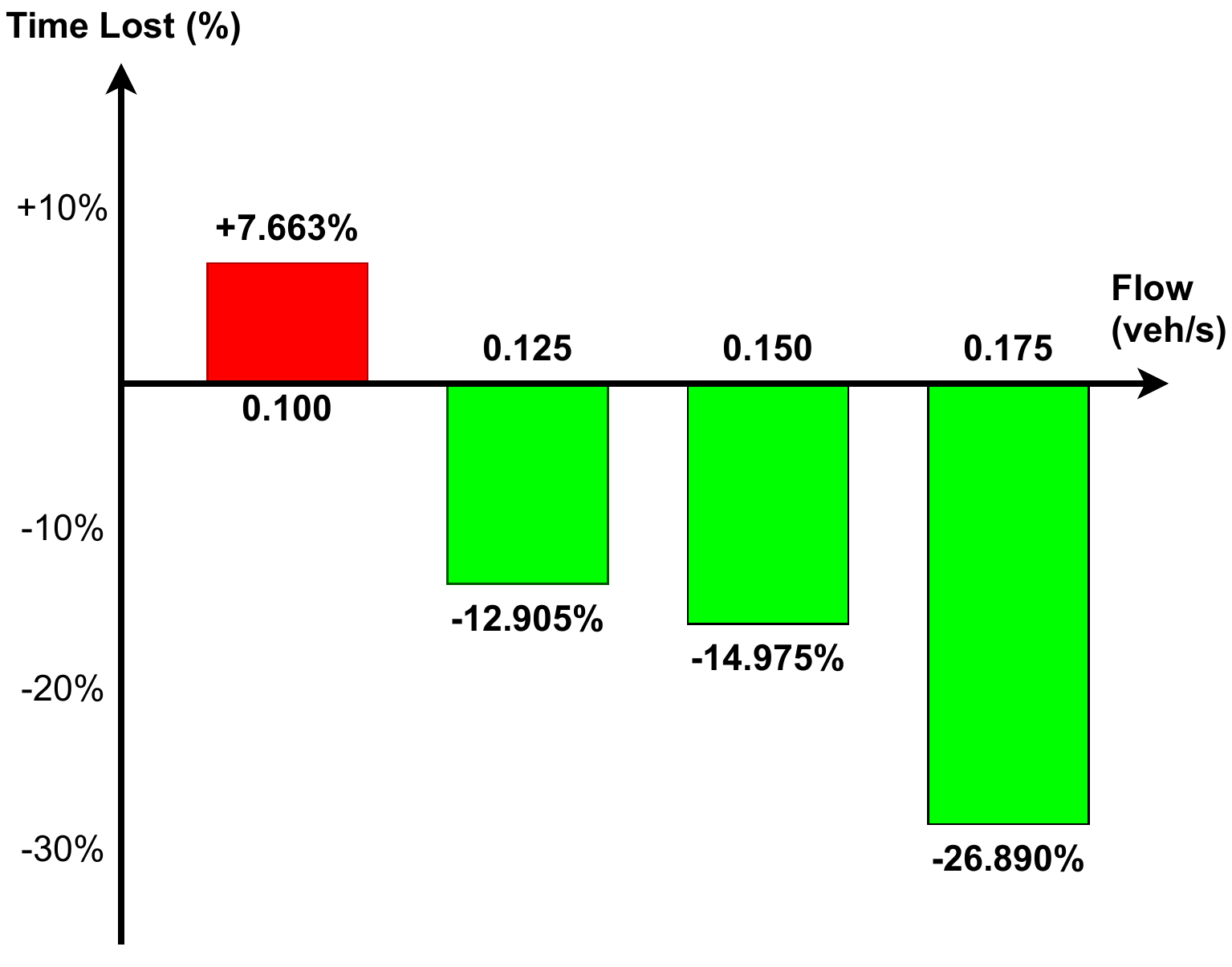}}
    \caption{The Performance of the Intelligent Traffic Signal Control in Experiment One.}
    \label{performance}
\end{figure}

\begin{figure}[htbp]
    \centerline{\includegraphics[scale=0.4]{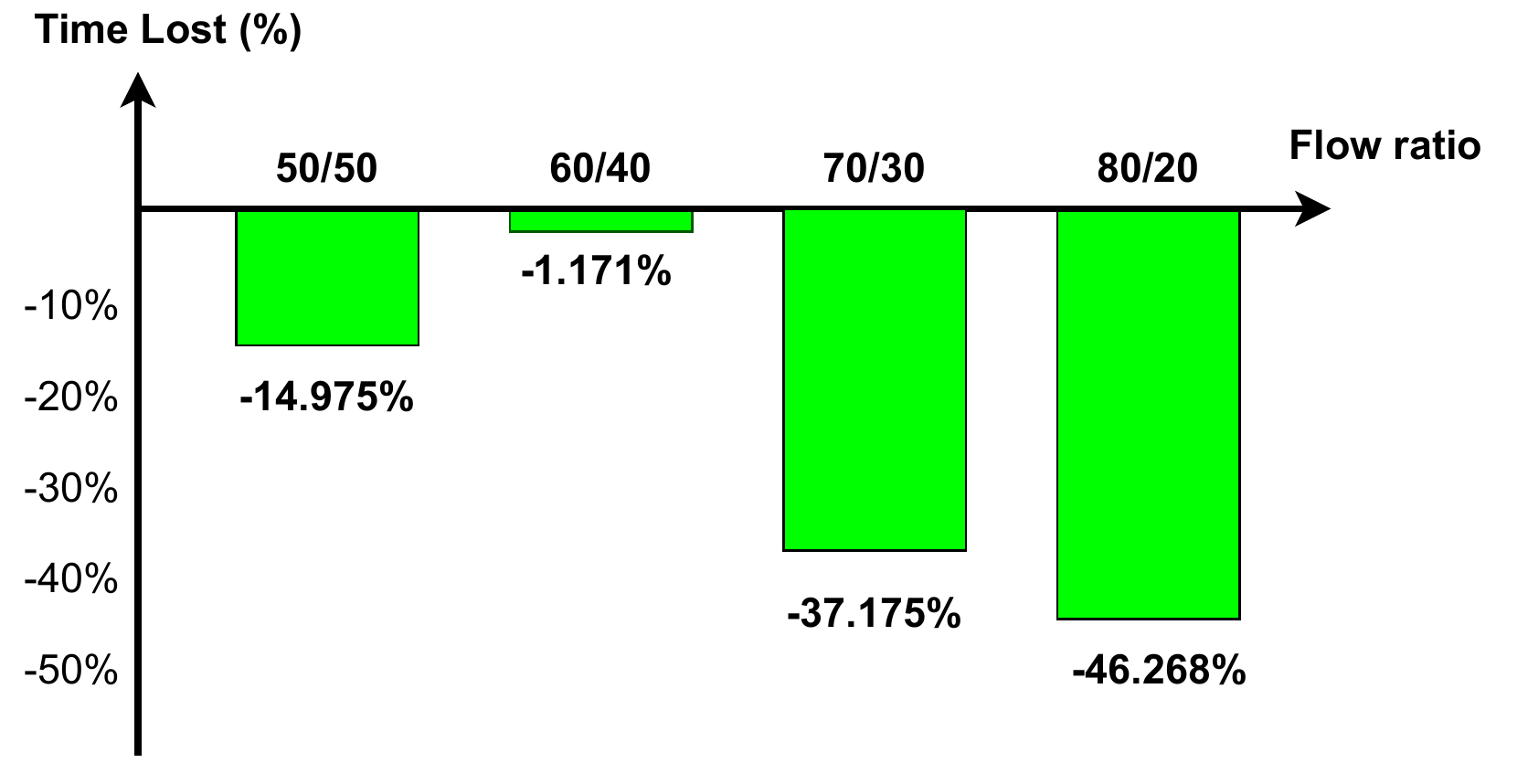}}
    \caption{The Performance of the Intelligent Traffic Signal Control in Experiment Two.}
    \label{performance2}
\end{figure}

The performance comparison between the intelligent traffic signal control and the fixed-time signal control is shown in Fig.~\ref{performance} and Fig.~\ref{performance2}. In the first experiment, although the time lost slightly increased in the low traffic flow conditions, the performance is excellent in the medium and high traffic flow situations. In the heaviest traffic condition, intelligent traffic signal control can almost save 27\% of the time which spends in the intersections. As for the major/minor road traffic conditions while the flow ratio is high. In the traffic flow ratio is at 80/20, the intelligent traffic signal control can almost save 50\% of the time which spends in the intersections. 

Because the control model of this system uses the traffic flow ratio of each road to decide the green phase time length, this paper confirmed that in the traffic flow ratio between 51$\sim$73\%, the intelligent traffic signal control could provide a good result.

\section{Conclusion}

In this paper, we apply Webster's method for controlling the traffic lights to relieve the traffic more effectively. As for the inputs for this method, we use object detection and multiple object tracking to make a vehicle counter, use the traffic flow model to get the maximum traffic flow of the road and develop a plan to calculate the real-time traffic flow. Moreover, we designed and built the roads for the experiment by using Virtual Test Drive software.

The simulation result proves this method can significantly decrease the time lost caused by the traffic signals and shorten the travel time by vehicles, not only for the equal flow ratio of each road but also the major/minor roads situations. The method can be used for other different traffic situations and can also foresee good results.

This system can be replaced by different control methods to achieve better performance for relieving the traffic or dealing with different types of traffic situations besides equal flow and major/minor road situations. Because of the use of cameras as the hardware foundation, the images can be used as different ways to get even more helpful information to help modern control methods to make decisions or respond to different emergencies on the road.

\section*{Acknowledgment}

Professor Chen has carefully guided this topic in the topic selection and research process. Professor Chen inquires about the research progress several times a week, gives us some tips, helps us develop research ideas, warmly encourages and carefully advises. Professor Chen's unconventional style, rigorous and realistic attitude and spirit of enlightenment teach us writing and study research. Although it only lasted a year, he has given us endless ways to be useful throughout our lives. The gratitude to Professor Chen cannot be expressed in words.

We want to express our gratitude to the members of the Automotive Electronics Lab., National Taipei University, Taiwan, for their passionate support and technical assistance during this project course. We also extend our acknowledgment to the MSC Software for their kindly supporting of the VTD tool.

From the beginning of the selection to complete the topic, how many respectable teachers, peers, and friends have given us powerful help. Please accept our sincere thanks!

\end{document}